\documentclass[traditabstract]{aa} 
\usepackage{graphicx}
\usepackage{txfonts,natbib}
\begin{document}

\title{The enigmatic central star of the planetary nebula PRTM~1\thanks{Based on observations made with the Very Large Telescope at Paranal Observatory under programmes 088.D-0750(B) and 078.D-0033(A).}}

   \author{
   Henri M. J. Boffin
   \inst{1}
   \and
   Brent Miszalski
   \inst{2,3}
   \and
   David Jones
   \inst{1}
          }

\institute{
European Southern Observatory, Alonso de Cordova 3107, Casilla 19001, Santiago, Chile\\
\email{hboffin@eso.org; djones@eso.org}
\and
South African Astronomical Observatory, PO Box 9, Observatory, 7935, South Africa
\and
Southern African Large Telescope Foundation, PO Box 9, Observatory, 7935, South Africa\\
\email{brent@saao.ac.za}
         }
   \date{Received -; accepted -}

\abstract{
The central star of the planetary nebula PRTM~1 (\object{PN G243.8$-$37.1}) was previously found to be variable by M. Pe\~na and colleagues. As part of a larger programme aimed towards finding post common-envelope binary central stars we have monitored the central star of PRTM~1 spectroscopically and photometrically for signs of variability. Over a period of $\sim$3 months we find minimal radial velocity ($\le$10 km s$^{-1}$) and photometric ($\le0.2$ mag) variability. The data suggest a close binary nucleus can be ruled out at all but the lowest orbital inclinations, especially considering the spherical morphology of the nebula which we reveal for the first time. Although the current data strongly support the single star
hypothesis, the true nature of the central star of PRTM~1 remains enigmatic and will require further radial velocity monitoring at higher resolution to rule out a close binary. If in the odd case that it is a close binary, it would be the first such case in a spherical planetary nebula, in contradiction to current thinking. 
}
   
   \maketitle
   
   \section{Introduction}
   At least $17\pm5$\% of planetary nebulae (PNe) have post common-envelope (CE) central stars with orbital periods less than $\sim$1 day \citep{m09}. 
 After a long and protracted debate (Balick \& Frank 2002), binary
central stars are now emerging as the preferred mechanism to shape PNe
(Soker 2006; Nordhaus \& Blackman 2006; De Marco 2009 and references
therein), long after their predicted existence and subsequent
discovery in the mid-70Õs (Paczynski 1976; Bond, Liller \& Mannery
1978).
 In order to understand the full influence of binaries on the formation and evolution of PNe a large and representative sample of binaries must be identified. Since the discovery of UU Sge as an eclipsing nucleus of the planetary nebula Abell 63 (Bond 1976, Bond et al. 1978), there are at least 40 close binary central stars now known, the overwhelming majority of which were found via photometric monitoring to reveal periodic variations due to irradiation, ellipsoidal modulation and eclipses (e.g. Miszalski et al. 2008, 2009a, 2011a, 2011b, 2011c; Corradi et al. 2011; Santander-Garc\'ia et al. 2012, in prep.). Many of these show a strong irradiation effect, due to the atmosphere of a K- or M-type dwarf companion being heated by the white dwarf primary, an effect that becomes undetectable once the orbital period exceeds a few days (De Marco, Hillwig \& Smith 2008). A strong selection effect therefore biases our current knowledge of binary central stars which may reside within a much larger fraction of the PN population (Moe \& De Marco 2006, 2012).

  \begin{figure*}[hbt]
      \begin{center} 
         \includegraphics[scale=0.45,angle=0]{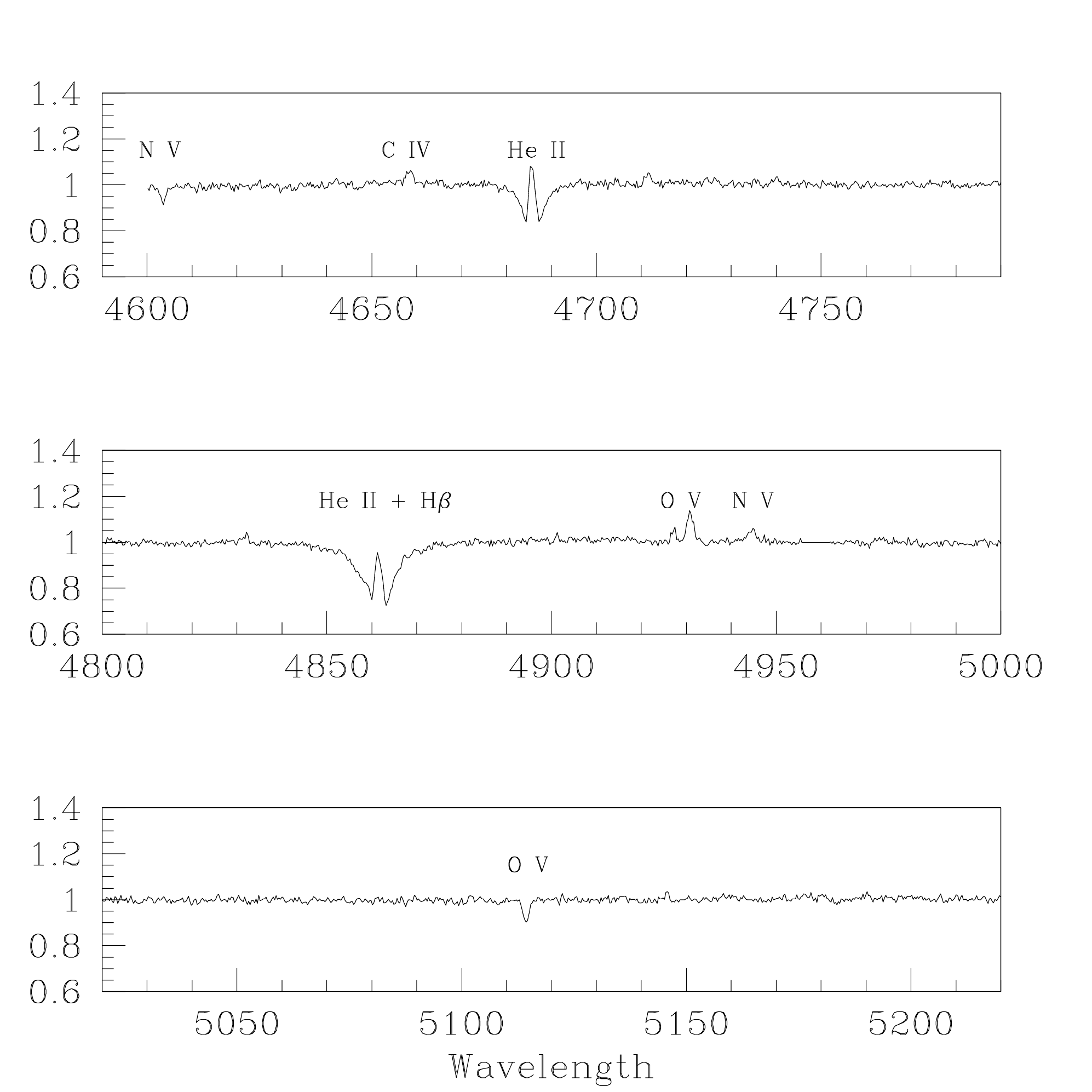}
         \includegraphics[scale=0.45,angle=0]{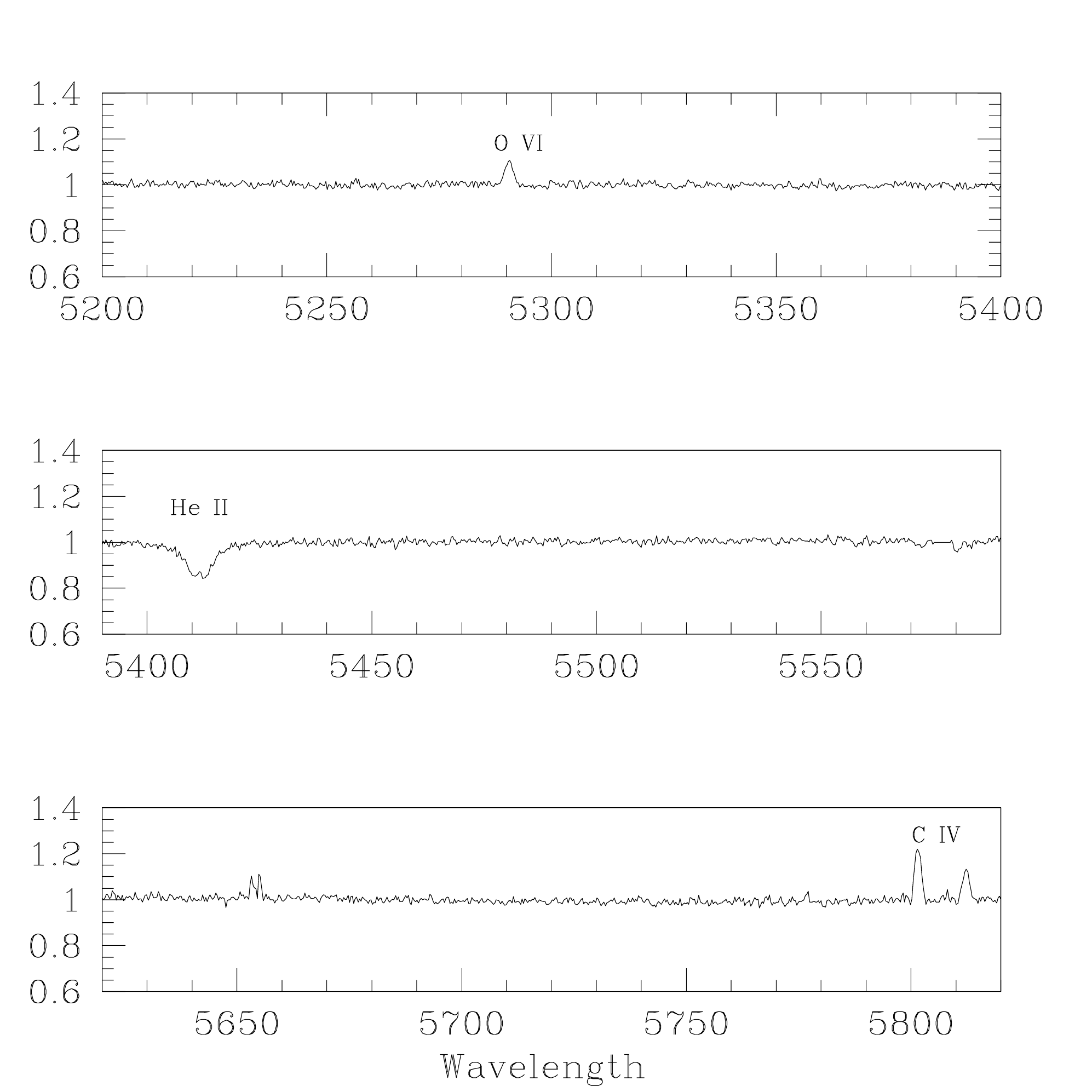} 
      \caption{Rectified average FORS2 spectrum of the central star of PRTM1. The nebular contribution has been mostly removed.}
        \label{fig:cspn}
         \end{center}
   \end{figure*}

   Radial velocity (RV) monitoring of central stars is an under-utilised methodology which has the capability to detect binaries that lack a favourable combination of orbital period and inclination to show detectable photometric variations. Boffin et al. (2012) powerfully demonstrated the potential of RV monitoring by discovering the double-degenerate nucleus of \object{Fleming~1} which showed a flat lightcurve and no irradiated emission lines (e.g. Miszalski et al. 2011b, 2011c). Previous RV monitoring surveys have unfortunately not been as successful with no routine discoveries made so far. M\'endez (1989) estimated 15\% of his monitored sample to show RV changes but none were periodic. High levels of variability were also found by De Marco et al. (2004), Sorensen \& Pollacco (2004) and Af{\v s}ar \& Bond (2005), but again no strictly periodic binaries were found (see also Frew et al. 2010). Although this could in principle partly be due to intrinsic, i.e. wind, variability, this can perhaps be mostly explained by the sparse sampling employed in these surveys, whereby spectra were often taken days or months apart, severely reducing their sensitivity to short periods. Improved sampling as in the Boffin et al. (2012) study is paramount to improving the utility of RV monitoring of central stars.

   The selection of Fleming~1 for RV monitoring by Boffin et al. (2012) was based on the extraordinary point-symmetric jets of Fleming~1 (e.g. L\'opez et al. 1993) and the unusual spectroscopic type of its central star. Its spectroscopic appearance matched that of the variable central star of the PN PRTM~1 (PN G243.8$-$37.1, Pe\~na et al. 1990 and Pe\~na \& Ruiz 1998, hereafter PRTM90 and PR98, respectively). If this unusual spectral type were to be a signature of a close binary nucleus, then PRTM~1 may also be expected to have one. Indeed, this possibility was raised by PR98 to explain the spectroscopic and photometric variability of PRTM~1. Following the successful monitoring of Fleming~1, we decided to test this hypothesis with time-series spectroscopy of the central star of PRTM~1 over 3 months. Here we report on the results of our campaign and present the first morphological study of the surrounding nebula. 

   This paper is structured as follows. Section~\ref{sec:prev} presents a very brief summary of previous work on this remarkable object. Section~ \ref{sec:obs} presents our spectroscopic and photometric study of the central star using the FORS2 instrument at the VLT. Section~\ref{sec:nebula} describes our imaging and kinematics of the nebula. A discussion of our results is in Sect.~\ref{sec:discussion} and Sect.~\ref{sec:end} concludes.

   \section{Previous work on PRTM~1}
   \label{sec:prev}
   Both the central star and nebula of PRTM~1 have been studied previously by numerous authors. 
   PRTM~1 was discovered by PRTM90 from objective prism survey plates and resembles other highly ionised Halo PNe (e.g. \object{NGC~4361}). The nebula spectrum is highly ionised and demonstrates only a slightly sub-solar abundance pattern, unusual for a Halo PN at a distance of $5\pm2$ kpc (PRTM90). PR98 reported the central star to have an O(H) spectral type with $T_\mathrm{eff}=80$ kK and log $g$=5.2 based on analysis of the central star by M\'endez (1991).\footnote{M\'endez (1991) referenced M\'endez et al. (1988) for this analysis. However, we did not find PRTM~1 to be included in this work.} PR98 found variability in the central star based on low-resolution spectroscopy which shows a blue continuum, deep H I and He II absorption lines, as well as faint and narrow C~IV $\lambda\lambda$5801, 5812 emission lines. The lack of C~III $\lambda$5696 emission rules out a [WC] classification, while non-detection of O~VI $\lambda\lambda$3811--34 emission rules out a [WO] classification (Crowther, De Marco \& Barlow 1998). 
   
   Feibelman (1998) found that \emph{IUE} observations obtained in 1988 and in 1990 showed significant differences, the stellar flux increasing by a factor of $\sim$3, while the He II $\lambda$1640 emission line was 1.5 times higher. PR98 also noticed important changes in the spectra of the central star, both in the continuum and absorption line profiles, when comparing spectra obtained on 28 October 1990, 1 February 1998 and 22 March 1998. Moreover, PR98 noted large variations in the stellar flux of the PN, with three deep minima that occurred on 20--22 November 1988, 10 August 1990, and 28 October 1990. The largest difference in brightness corresponded to $\Delta V$$\sim $1 mag. PR98 concluded that the stellar continuum changed by a factor 2--3 on short timescales (weeks) and 20\%--30\% variations occurring within days. 
   
   PR98 proposed 3 possible scenarios to explain these variations: (a) a close binary system with a period of a few hours to a few days that shows eclipses or an irradiation effect, (b) non-radial pulsations in a very hot high-gravity pre-white dwarf, or (c) stellar wind variations. Our observational campaign aimed to test their preferred scenario of a close binary central star (a).

   \section{The central star}
   \label{sec:obs}
   PRTM~1 was observed with the FORS2 instrument (Appenzeller et al. 1998) under the Very Large Telescope (VLT) service mode programme 088.D-0750(B). Table~\ref{tab:rv} lists the 15 useful spectra taken over the course of 84 days from 11 October 2011 to 3 January 2012. A 0.5\arcsec\ slit and the 1400V volume-phase holographic grating were used to give a wavelength range of 4597--5901 \AA\ with a dispersion of 0.32 \AA\ pixel$^{-1}$ and a spectral resolution measured from the arc spectrum of 1.22 \AA\ (full width at half-maximum, FWHM). Exposure times mostly varied between 900--1500 s resulting in an average signal-to-noise ratio (S/N) of 105 near 5100 \AA. Basic reductions of the data were performed using the ESO FORS pipeline and one-dimensional spectra were optimally extracted using \textsc{midas}. The mean spectrum of the central star of PRTM1 is shown in Fig.~\ref{fig:cspn}.
   Since it is not possible to take contemporaneous arc exposures with FORS2, the error bar on the heliocentric radial velocity is rather high, and  we therefore do not attempt at measuring it (but see Sect.~4.2).
   
   In view of the results of PR98 who found large variations in the spectroscopic behaviour of the central star, we compared our higher resolution spectra against theirs. No major changes were seen over the full 84 day monitoring period. We also measured the equivalent width of the various absorption lines and did not see any significant change. The absorption line at 5114.06  \AA, most likely due to O~V, has a rather constant equivalent width of 85~m\AA, the $1\sigma$ variation being of 7.7~m\AA, i.e. a change of $\sim$9\%, well below the factor 2--3 found by PR98.
    
   \begin{table}
      \caption{Spectroscopic observations and measured radial velocities of the central star of PRTM~1. The table shows, respectively, the modified julian date (MJD) at the midpoint of each observation, the exposure time, the attained signal-to-noise ratio and the measured radial velocity (see text). We estimate a typical error of 3 km s$^{-1}$ on the latter.}
      \label{tab:rv}
      \centering
      \begin{tabular}{lrrr}
         \hline \hline
         MJD = JD-2400000.5 & Exposure & S/N                & Radial velocity\\
                 &  (s)    & ($\lambda=5100$\AA) &  (km s$^{-1}$) \\
                 \hline
               55845.199 & 540 & 50 & $-$11.8\\  
               55856.281 & 1200 & 85 & $-$13.1\\
               55857.354 & 1200 & 140 & $-$4.2\\
               55858.218 & 1200 & 95 & $-$9.5\\
               55875.158 & 1200 & 75 & 2.1\\
               55877.131 & 1500 & 90 & $-$10.1\\
               55879.254 & 1350 & 105 & $-$5.0\\
               55880.136 & 1500 & 110 & $-$14.9\\
               55882.175 & 1200 & 110 & $-$0.2\\
               55901.531 & 1200 & 110 & 2.8\\
               55903.279 & 1500 & 95 & 6.0\\
               55905.270 & 1200 & 110 & 0.3\\
               55906.063 & 1050 & 130 & $-$6.1\\
               55920.147 & 900 & 50 & 3.2\\
               55929.124 & 900 & 130 & 2.5\\
\hline
      \end{tabular}
   \end{table}

   \begin{figure}
      \begin{center}
        \includegraphics[scale=0.35,angle=270]{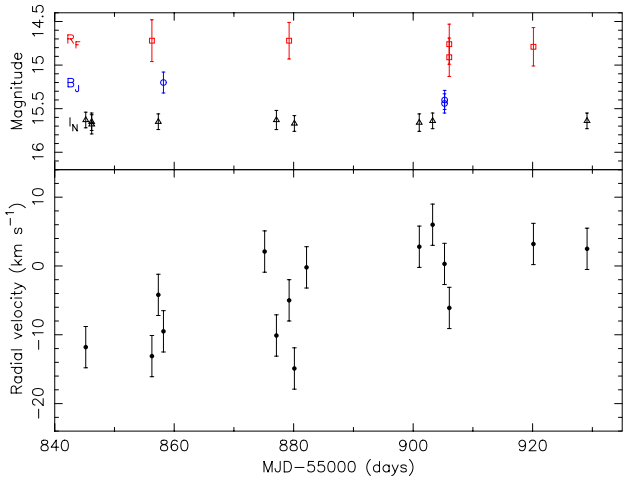}
      \end{center}
      \caption{Photometric and spectroscopic measurements of PRTM1 during our 84-day campaign. The top panel shows the estimated $R_F$, $B_J$, and $I_N$ magnitudes as a function of time, while the bottom panel shows the radial velocity of the O~V 5114.06 absorption feature in the rest-frame of the nebula (see text).}
      \label{fig:series}
   \end{figure}

   The FORS2 spectra were converted to a logarithmic wavelength scale before shifting each spectrum to match the systemic nebula RV based on the strong [O~III] emission lines. Gaussian profile fitting of the nearby O~V $\lambda$5114.06 absorption line was used to measure the central star's RV with respect to the nebula RV. Table~ \ref{tab:rv} lists the derived RVs together with the S/N reached. Similar RV variations were found from other stellar features (e.g. He~II $\lambda$5412, C~IV $\lambda$5801, $\lambda$5812) which all moved in phase with O~V. 
   From the signal-to-noise of the spectra and the spectral resolution, we estimate, from an extensive set of Monte Carlo simulations, an error of 3 km/s per data point -- or slightly smaller for the best of our spectra. This error encompasses both the (small) error related to the position of the nebula emission line and that related to the position of the O~V line. The slit used was  0.5 \arcsec wide and was therefore always smaller than the seeing.
   
   Figure~\ref{fig:series} shows the RV measurements over time. The RV is apparently changing with time, spread over a range from $-14.9$ to $+6.0$ km s$^{-1}$, with variations occurring on a time scale of a few days. The mean value is $-3.7$ km s$^{-1}$ and we measure an intrinsic scatter of 6.6 km s$^{-1}$, much larger than the individual errors. The figure seems to indicate a slow upward trend, but this may simply be due to the limited sampling. To quantify whether this trend is real, we have compared the residuals when removing or not the linear trend identified in Figure~\ref{fig:series}: the $\chi ^2$ changes from 47 to 29, and a $F$-test applied to these number reveals that this is not significant (probability = 0.38). We therefore conclude that this trend is most likely spurious.

   During the campaign a variety of acquisition images were also taken to monitor any photometric variability and to study the morphology of the nebula (Sect. \ref{sec:nebula}). Table \ref{tab:acq} gives the full list of acquisition images taken. The observations were performed using the filters OII+44 (central wavelength, $\lambda _c=371.7$ nm, width, $W_o=7.3$ nm), OIII+50 ($\lambda _c=500.1$ nm, $W_o=5.7$ nm), H\_Alpha+83 ($\lambda _c=656.3$ nm, $W_o=6.1$ nm) and I\_BESS+77 ($\lambda _c=768.0$ nm, $W_o=138.0$ nm). An approximate absolute scale to our photometry was determined by using the catalogue photometry of the SuperCOSMOS Sky Survey (Hambly et al. 2001), namely $I_N$ for I\_BESS+77, $R_F$ for H\_Alpha+83 and $B_J$ for OIII+50. Differential photometry was performed using more than 10 field stars in all filters, using the same methodology as that used by Miszalski et al.~(2011c), i.e. using {\tt sextractor} and an aperture  of radius of 3$\arcsec$ ($\sim$1.5 the worse seeing). We also made an analysis using standard aperture photometry and obtained similar results. There were insufficient stars in the [O~II] frames to derive an appropriate calibration. Note that because of the large aperture used, our values partly include a nebula contribution (affecting the $B_J$ and $R_F$ magnitudes the most). These magnitudes are included in Table \ref{tab:acq} and displayed alongside the RVs in Fig. \ref{fig:series}. There is no evidence for a change in central star brightness exceeding $\sim$0.2 mag during the time of our spectroscopic monitoring. 

   \begin{table*}
      \centering
      \caption{Log of acquisition images and derived magnitudes for PRTM~1. The magnitudes include nebula and central star in the aperture.}
      \label{tab:acq}
      \begin{tabular}{llrrrrr}
         \hline \hline
         MJD & Filter & Exptime & $B_J$ & $R_F$ & $I_N$ & Seeing\\
             &        & (s)     &       &       &       & (FWHM, arcsec)\\  
         \hline
55845.189142& I\_BESS+77	&    10	&	-  	&	-	     & 15.63 &	1.09\\
55846.181415& I\_BESS+77	&    10	&	-	   &	-	     & 15.65 &  2.04\\
55846.182760& I\_BESS+77	&    10	&	-	   &	-	     & 15.68 &  2.12\\
55856.263531& H\_Alpha+83	&    60	&	-	   &	14.72  & -	    &  1.33\\
55857.336515& I\_BESS+77	&    10	&	-	   &	-	     & 15.65 &  0.68\\
55858.201558& OIII+50	   &    60	&	15.20	&	-	     & -	    &  1.05\\
55875.139573& OII+44	      &    60	&	-     &	-	     & -	    &  1.41\\
55877.110056& I\_BESS+77	&    10	&	-	   &	-	     & 15.63 &  0.81\\
55879.234569& H\_Alpha+83	&    90	&	-	   &	14.72  & -	    &  1.12\\
55880.115462& I\_BESS+77	&    30	&	-	   &	-	     & 15.67 &  0.89\\
55882.157167& OII+44	   &    60	&	-	   &	-	     & -     &  0.90\\
55901.023047& I\_BESS+77	&    10	&	-	   &	-	     & 15.66 &  1.10\\
55903.258718& I\_BESS+77	&    10	&	-	   &	-	     & 15.64 &  0.98\\
55905.250492& OIII+50  	   &    60	&	15.44	&	-	     & -     &  0.57\\
55905.272043& OIII+50	   &   120 & 15.40	& -	     & -     &  0.57\\
55906.045124& H\_Alpha+83	&    30	&	-	   &	14.76  & -     &  0.46\\
55906.046464& H\_Alpha+83	&    60	&	-	   &	14.91  & -     &  0.50\\
55920.133308& H\_Alpha+83	&    60	&	-	   &	14.79  & -     &  0.53\\
55929.105362& I\_BESS+77	&    10	&	-	   &	-	     & 15.64 &  0.48 \\
         \hline
      \end{tabular}
   \end{table*}

   \begin{figure*}
      \begin{center}
        \includegraphics[scale=0.4]{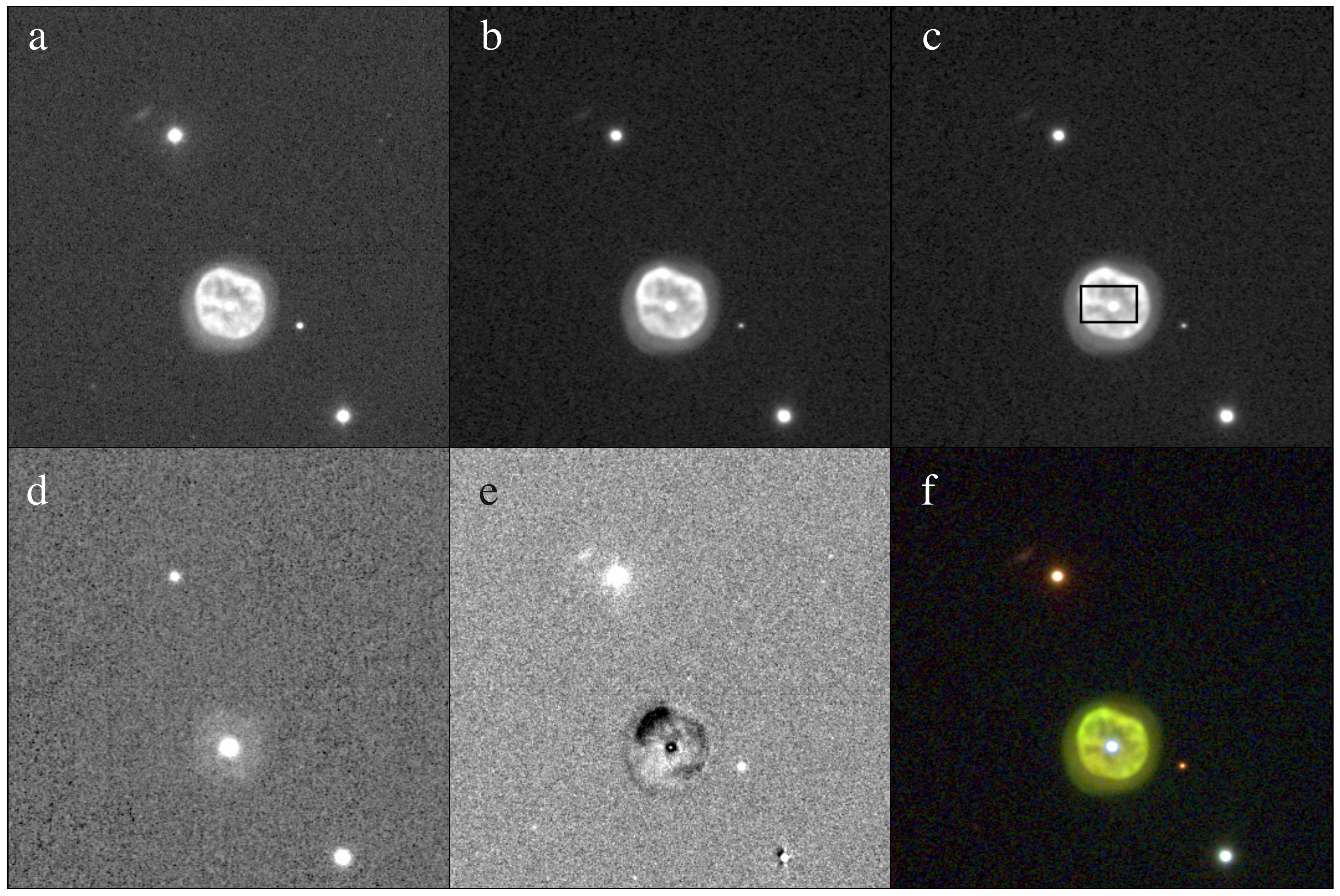}
      \end{center}
      \caption{VLT FORS2 images of PRTM1. (a) H$\alpha$, (b) [O~III], (c) [O~III] with the ARGUS $11.5\times7.3$ arcsec$^2$ footprint (see Sect. \ref{sec:kin}), (d) [O~II], (e) H$\alpha$ divided by [O~III] (H$\alpha$ excess appears light and [O~III] dark), and (f) Colour-composite made from H$\alpha$ (red), [O~III] (green) and [O~II] (blue). The images measure $90\times90$ arcsec$^2$ with North up and East to left.}
      \label{fig:montage}
   \end{figure*}

   \section{The nebula}
     \label{sec:nebula}
   \subsection{Morphology}
\begin{figure}[h]
      \begin{center}
        \includegraphics[scale=8.0]{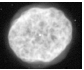}
      \end{center}
      \caption{NOAO H$\alpha$ image of NGC~1501 (Credit: Jay Gallagher U. Wisconsin/WIYN/NOAO/NSF). The diameter of the nebula is $\sim$60\arcsec\ across and North is up with East to left. Note the strong resemblance to PRTM~1 (Fig. \ref{fig:montage}).}
      \label{fig:n1501}
   \end{figure}

   Figure \ref{fig:montage} shows the first high-resolution images taken of PRTM~1 featuring the observations with the best seeing in Table \ref{tab:acq}.\footnote{Specifically data taken on the nights of MJD=55882 ([O~II]), MJD=55905 ([O~III]) and MJD=55906, 55920 (H$\alpha$).} The data show a filamentary inner nebula 14--15\arcsec\ across surrounded by a fainter shell measuring $20.5\times21.3$ arcsec. At the edges of the apparent major axis along PA$\sim$135$^\circ$ there seems to be two thin arcs in the [O~III] image. The nebula is typical of multiple-shell PNe (Chu, Jacoby \& Arendt 1987) and the somewhat `dented' appearance of the inner nebula is unusual but also seen in other PNe (e.g. \object{NGC~2867}, Schwarz, Corradi \& Melnick 1992). Given the very high degree of nebula ionisation there are no low-ionisation structures present and no faint features located outside the fainter shell. Together all these properties are inconsistent with the nebula morphologies of post common-envelope PNe (Miszalski et al. 2009b). It is interesting to note that several of the nebulae surrounding GW Vir pulsating central stars strongly resemble PRTM~1 including NGC~7094, Abell~43, NGC~2867 and especially \object{NGC~1501} (Fig. \ref{fig:n1501}). 
   
   \subsection{Kinematics}
   \label{sec:kin}
   To study the kinematics of the nebula of PRTM~1, we retrieved high resolution observations from the ESO archive. The PN was observed on 15 January 2007 using the FLAMES facility of the VLT (Pasquini et al. 2002) under programme 078.D-0033(A) (PI: D. Sch\"onberner, see Sch\"onberner et al. 2010). FLAMES was operated in its ARGUS mode, namely an integral field unit (IFU) of $22\times14$ microlenses, each 0.52\arcsec\ a side, giving a footprint of $11.5\times7.3$ arcsec$^2$ (see Fig. \ref{fig:montage}c). The HR8 grating was selected for the GIRAFFE spectrograph providing a resolving power $R=32,000$ around the [O~III] $\lambda$5007 emission line. Two 1800-s exposures were retrieved from the ESO archive and reduced using the ESO pipeline before being averaged using standard \textsc{starlink} routines. The seeing during the observations varied between 1.0--1.5\arcsec.

\begin{figure}
      \begin{center}
        \includegraphics[scale=0.35,angle=270]{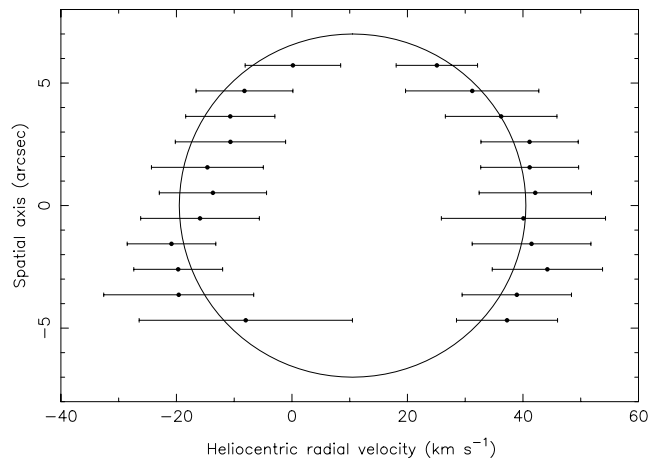}
      \end{center}
      \caption{VLT ARGUS velocity profile of PRTM~1 from a 1.04\arcsec\ longslit trace, overlaid with the profile expected for a spherical nebula at a heliocentric systemic velocity of 12.5 km s$^{-1}$. Note that for clarity and  ease of comparison to a spherical model only one PA is presented, but the kinematics were examined across the entire nebula and are entirely consistent with the PA presented.}
      \label{fig:kin}
   \end{figure}

   Figure \ref{fig:kin} shows Gaussian fits to an extracted spectrum equivalent to a 2 spaxel-wide (1.04\arcsec) longslit spectrum through the CSPN at a position angle of 90\degr{}. The velocity profile shows little deviation from that expected from a spherical nebula, strongly indicating that PRTM~1 is roughly spherical with an expansion velocity of $30.0\pm5.0$ km~s$^{-1}$ and a heliocentric systemic radial velocity of $12.5\pm5.0$ km~s$^{-1}$. The relatively large error bars reflect turbulence in the nebula. Assuming a distance of $5\pm2$ kpc (PRTM90), the kinematic age of the $\sim$0.5 pc diameter nebula corresponds to $5500^{+3800}_{-2700}$ years including both errors in the expansion velocity and distance. Both the kinematics and morphology of PRTM~1 are consistent with those of NGC~1501 (Sabbadin \& Hamzaoglu 1982). 

  \begin{figure}[hbt]
      \begin{center} 
        \includegraphics[scale=0.45,angle=0]{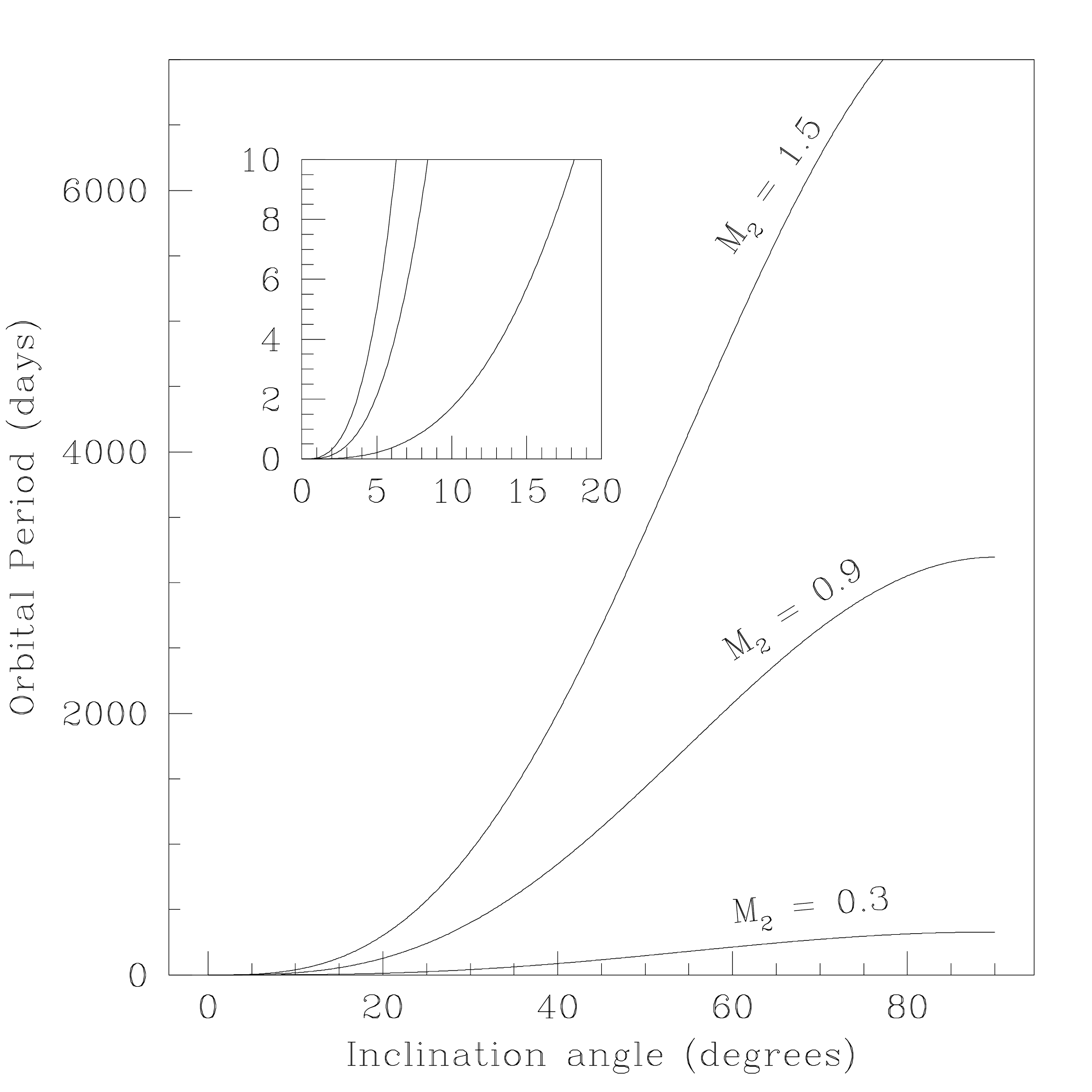} 
      
      \caption{Possible values of the orbital period as a function of the inclination angle for 3 values of the secondary mass. The inset shows a zoom for the smallest inclination angles and periods.}
        \label{fig:semi}
        \end{center}
   \end{figure}

\section{Discussion}
\label{sec:discussion}
In combination with the observations of Pe\~na and colleagues, the observations presented in this paper allow us to refine the possible nature of the central star. We discuss three possible scenarios: a post common-envelope binary, a wider binary with an orbital period of weeks or months to years and a single central star. 

\subsection{A close binary central star?}
A thorough periodogram search was performed, both on the observed RV time-series and the corrected series where we removed the spurious long-term trend. Although some periods phased up reasonably well (e.g. $P$$\sim$3 hours), these always had at least 2 uncooperative RVs that could not be explained away --  i.e. they were more than 5-$\sigma$ outliers. In any case, these periods had too low a S/N in the periodogram, meaning that no clean periods were found. 
If the RV variations we have observed were due to the presence of a companion, the orbital period would have to be smaller than around a few days, to account for the wide changes seen in this time range. The semi-amplitude we measure would be on the order of 10 km~s$^{-1}$, which given reasonable values for the masses of the components would mean that the system is seen almost face-on. We can in fact try to be more quantitative. Using the definition of the RV semi-amplitude in a circular Keplerian orbit, we can define a relation between the orbital period and the inclination of the system in the plane of the sky, assuming the masses of the components. This is shown in Fig.~\ref{fig:semi}, where we have taken 0.6~M$_\odot$ as the mass of the white dwarf, and considered companions of masses 0.3, 0.9 and 1.5~M$_\odot$, respectively. The last value is probably unrealistically too high as PRTM1 is a Halo PN, whose progenitor must have had an initial mass well below this value and there is no reason to think that the secondary became more massive than the primary. Not surprisingly given the very low semi-amplitude measured, for most of the values of the inclination angles, the orbital period would have to be rather large, above 100 d (for M$_2$=0.3~M$_\odot$) or even 1000 d (for M$_2$=1.5~M$_\odot$). As mentioned above, the variations we detect are on a much smaller time scale than those detected by PR98 and Feibelman (1998) and therefore could not be due to such a long period. Only if the inclination angle were very small would the orbital period be below a few days for the whole range of masses. The almost perfectly round shape of the nebula unfortunately does not provide any clue on the possible presence of a torus or ring, and thus no clue to obtain the inclination of the orbital plane if it exists (see e.g.  Jones et al. 2012; Tyndall et al. 2012).

None of the spectra obtained here or in PR98 show emission lines of C~III and N~III that would come from an irradiated atmosphere of a main-sequence companion (e.g. Miszalski et al. 2011c). Nor do we find any spectral features that can be attributed to a companion, all stellar features move in lockstep with each other. The roughly spherical nebula morphology and kinematics (Sect.~\ref{sec:nebula}) are also incompatible with a close binary on theoretical grounds (e.g. Soker 1997) and the observed morphologies of nebulae around currently known close binaries (e.g. Miszalski et al. 2009b; Miszalski 2012). While we cannot rule out a very low inclination close binary based on the relatively small number of observations currently in hand, it seems unlikely that future observations would find one. If there were a close companion it would probably be a low mass white dwarf.

\subsection{A wide binary central star?}
If the period were weeks to months we may have expected to see more coherent radial velocity variations over the $\sim$3 months we observed the central star. Moreover, the relatively large level of spread in the RVs of $\sigma\sim$10 km/s happening on a timescale of a few days appears to be real. If some larger periodicity would have to be invoked, it would have to be {\em in addition} to some other intrinsic variation. As we indicated above, the RV data shows  some {\it apparent} upward trend. If we discard the possibility that this is simply due to small number statistics and our sampling -- as we have shown above that it probably is -- we may infer that the mean velocity increased by 15.7 km~s$^{-1}$ over the 84-day period. Thus the orbital period in this case would be more than twice this, and the radial velocity amplitude should be of the order of 7--10 km s$^{-1}$. The longer the period, the larger the radial velocity amplitude would have to be. As such amplitude is, however, inversely proportional to a power of the period, too large a period would be impossible. With our values, we therefore estimate that the largest possible period would be of $\sim$150--200 days.

\subsection{A single central star?}
If PRTM~1 were to have a single central star, then we would have to explain the apparent RV variations as due to either pulsations or perhaps wind variability. We can probably rule out non-radial GW Vir type pulsations since the log $g$ of 5.2 is $\sim$ 2.0 dex too low for the $T_\mathrm{eff}=80$ kK temperature to place it in the expected instability domain (Fontaine \& Brassard 2008). Furthermore, the O(H) spectral type of the central star of PRTM~1 is inconsistent with a pulsator since it cannot be classified as a PG1159 (Napiwotzki \& Sch\"onberner 1995) or an early-[WC] central star (Crowther et al. 1998). Both of these spectral types dominate the sample of known GW Vir pulsating central stars (e.g. Ciardullo \& Bond 1996). Wind variability would only be expected if PRTM~1 were an Of(H) type central star (e.g. M\'endez, Herrero \& Manchado 1990). However, the spectrum clearly does not have this spectral type (Fig. \ref{fig:cspn}). Another possibility is that secular evolutionary changes could explain the small amount of photometric variability constrained by our observations and the previously observed high levels of variability (see Sect. 1), as seen in some objects but on a much smaller scale (e.g. Bond 2008). Such changes may be one possible explanation as to why we find no variability compared to previous authors.
Finally, we re-emphasise the spherical morphology of the nebula that is consistent with a single central star.

\section{Conclusions}
\label{sec:end}
We have presented in this paper a new study of the nebula and central star of PRTM~1. Our imaging reveals a rather spherical morphology, and based on archival data, the kinematics are also consistent with a spherical nebula. We also conducted a spectroscopic monitoring campaign of the central star of PRTM~1. Unlike previous authors, we do not find large variations in the continuum or spectral features over the full time range of our observations (84 days). We notice, however, some radial velocity variations of the CSPN but were unable to find any coherent periodicity. At this stage,  it is not possible to
decide whether the CSPN is a binary or not, however the binary hypothesis seems very
unlikely.
If PRTM~1 were indeed a close binary, then it would be the first one discovered within a spherical nebula in contradiction with our understanding of how close binaries should shape PNe (e.g. Soker 1997). 

\begin{acknowledgements}
   BM thanks ESO Chile for their hospitality and the opportunity to participate in their visitor programme during January 2012. This work was co-funded under the Marie Curie Actions of the European Commission (FP7-COFUND). It is a pleasure to thank Ralf Napiwotzki and an anonymous referee for useful comments and suggestions that improved the paper.
\end{acknowledgements}

\end{document}